# The Green-function transform and wave propagation

**Colin J. R. Sheppard,[1] S. S. Kou[2] and J. Lin[2]**

[1]Department of Nanophysics, Istituto Italiano di Tecnologia, Genova 16163, Italy

[2]School of Physics, University of Melbourne, Victoria 3010, Australia

*Corresponding author: colinjrsheppard@gmail.com

Abstract: Fourier methods well known in signal processing are applied to three-dimensional wave propagation problems. The Fourier transform of the Green function, when written explicitly in terms of a real-valued spatial frequency, consists of homogeneous and inhomogeneous components. Both parts are necessary to result in a pure out-going wave that satisfies causality. The homogeneous component consists only of propagating waves, but the inhomogeneous component contains both evanescent and propagating terms. Thus we make a distinction between inhomogenous waves and evanescent waves. The evanescent component is completely contained in the region of the inhomogeneous component outside the *k*-space sphere. Further, propagating waves in the Weyl expansion contain both homogeneous and inhomogeneous components. The connection between the Whittaker and Weyl expansions is discussed. A list of relevant spherically symmetric Fourier transforms is given.

## 1. Introduction

In an impressive recent paper, Schmalz *et al.* presented a rigorous derivation of the general Green function of the Helmholtz equation based on three-dimensional (3D) Fourier transformation, and then found a unique solution for the case of a source (Schmalz, Schmalz et al. 2010). Their approach is based on the use of generalized functions and the causal nature of the out-going Green function. Actually, the basic principle of their method was described many years ago by Dirac (Dirac 1981), but has not been widely adopted. The present paper presents some of the important implications of this approach. An aim of the present paper is to reconcile the different approaches used in the signal processing and theoretical physics disciplines, respectively. We discuss the physical implications with respect to evanescent and propagating waves, in-going and out-going waves, and forward and backward propagating waves. Some self-consistent spherically symmetric Fourier transforms pairs are listed.

A simple source, equivalent to the Green function, impulse response, or point-spread function, is of fundamental importance in diffraction, wave propagation, optical signal processing, and so on, and has a Fourier transform that can be recognized as a transfer function. Recently, we have shown that in three dimensions there is some advantage, from the point of view of both computational utility and conceptual understanding, in introducing the 3D transform of the Green function (Lin, Yuan et al. 2011; Lin, Rodriguez-Herrera et al. 2012; Kou, Sheppard et al. 2013; Sheppard, Lin et al. 2013). Thus, unlike many other works we are particularly interested in the properties and significance of the 3D transform in its own right, rather than just as a step towards developing the real space Green function.

# 2. The Green-function transform

## 2.1 The homogeneous solution

We start by considering the homogeneous, scalar, time-independent Helmholtz equation in 3D empty, free space:

$$(\nabla^2 + k_0^2)U(\mathbf{r}) = 0, \tag{1}$$

where $k_0$ is the magnitude of the wave vector, $k_0 = 2\pi / \lambda$. As we all know, the general solution is

$$U(r) = A\frac{\sin k_0 r}{r} + B\frac{\cos k_0 r}{r}, \tag{2}$$

where $A, B$ are arbitrary constants, in general complex. The second term exhibits a singularity at $r = 0$, which is not an appropriate physical solution in free space, so we recognize the homogeneous solution as

$$U(r) = A\frac{\sin k_0 r}{r} = A\frac{\exp(ik_0 r)}{2ir} - A\frac{\exp(-ik_0 r)}{2ir}. \tag{3}$$

The homogeneous solution thus represents a combination of a source (an out-going wave) and a sink (also called a drain or outlet (Tyc and Zhang 2011), and corresponds to an in-going wave) in antiphase, respectively.

## 2.2 The inhomogeneous solution

The standard method of deriving the Green function, given in many physics or electromagnetism texts (Titchmarsh 1948; Jackson 1962; Morse and Feshbach 1978), is to Fourier transform the inhomogeneous Helmholtz equation, with a forcing term $-4\pi\delta(\mathbf{r} - \mathbf{r}_0)$,

$$(\nabla^2 + k_0^2)U(\mathbf{r}) = -4\pi\delta(\mathbf{r} - \mathbf{r}_0), \tag{4}$$

to give

$$(-k^2 + k_0^2)\tilde{U}(\mathbf{k}) = -4\pi e^{-i\mathbf{k}\cdot\mathbf{r}_0}, \tag{5}$$

so that

$$\tilde{U}(\mathbf{k}) = \frac{4\pi e^{i\mathbf{k}\cdot\mathbf{r}_0}}{(k^2 - k_0^2)}. \tag{6}$$

Then, inverse Fourier transforming using an appropriate contour integration, we can obtain different solutions:

$$\begin{aligned} \frac{4\pi e^{i\mathbf{k}\cdot\mathbf{r}_0}}{(k^2 - k_0^2)} &\Rightarrow \frac{\exp[ik_0(r - r_0)]}{r}, \\ \frac{4\pi e^{i\mathbf{k}\cdot\mathbf{r}_0}}{(k^2 - k_0^2)} &\Rightarrow \frac{\exp[-ik_0(r - r_0)]}{r}. \end{aligned} \tag{7}$$

By choice of the correct contour we can thus generate either an out-going or in-going wave. In many works, it is then claimed that as a consequence of the Sommerfeld radiation condition (Ausstrahlungsbedingung) (Sommerfeld 1949, p. 189), (Schot 1992), sinks do not occur naturally and thus we can select the source as giving the correct solution. After putting $\mathbf{r}_0 = \mathbf{0}$, we thus have (according to many sources (Titchmarsh 1948; Jackson 1962; Morse and Feshbach 1978), and also Mathematica),

$$\frac{4\pi}{(k^2 - k_0^2)} \Leftrightarrow \frac{\exp(ik_0 r)}{r}. \tag{8}$$

The Hankel transform is thus calculated using integration in the complex plane. While it is recognized that contour integration is a powerful method that can give an analytic solution leading to many useful results, it should be appreciated that integration along different paths will lead to different results in the presence of singularities, and so interpretation of a result for a particular path as being the Hankel transform is not always appropriate. An alternative view is based on the use of generalized functions, as usually done in the signal processing discipline. This approach can be put on a rigorous footing using distribution theory or generalized derivatives (Temple 1955; Lighthill 1958) as discussed by Schmalz *et al.* (Schmalz, Schmalz et al. 2010). Further, we can take advantage of the fact that Hankel transforms constitute a reciprocal, unique transform pair.

We can identify two issues with the standard treatment. First, it is not clear why the delta forcing term in Eq. 4 should represent a source rather than a sink, or a source/sink combination. Second, we note that dividing by $k^2 - k_0^2$ to obtain Eq. 6 is valid only for $k^2 \neq k_0^2$, so we should write (Lighthill 1958; Dirac 1981; Schmalz, Schmalz et al. 2010), adding an arbitrary multiple of the homogeneous solution which does not affect Eq. 5,

$$\begin{aligned} \frac{4\pi}{(k^2 - k_0^2)} + C\,\delta(k^2 - k_0^2) &\Rightarrow \frac{\cos k_0 r}{r} + i\frac{\sin k_0 r}{r}, \\ \frac{4\pi}{(k^2 - k_0^2)} + D\,\delta(k^2 - k_0^2) &\Rightarrow \frac{\cos k_0 r}{r} - i\frac{\sin k_0 r}{r}. \end{aligned} \tag{9}$$

where $C, D$ are different constants, in general complex. Thus we see that the choice of the arbitrary multiple can give the solution for either a source or a sink, or a combination. Subtracting the two Eqs. 9,

$$E\,\delta(k^2 - k_0^2) \Rightarrow \frac{\sin k_0 r}{r}, \tag{10}$$

where $E = i(D - C)/2$ is real, as the term on the right is purely real. This then implies, from Eq. 9, that

$$\frac{4\pi}{(k^2 - k_0^2)} \Rightarrow \cos(k_0 r)/r, \tag{11}$$

which represents the inhomogeneous solution, rather giving the relationship in Eq. 8.

### 2.3 The spherically symmetric Fourier transform

In this paper we will use the definition of Fourier transform with no premultiplying constants (Bracewell 1978), called in Mathematica the signal processing convention, so that the components of spatial frequency $\mathbf{q} \in \mathbb{R}^3$ are $(q_x, q_y, q_z)$, and $q = (q_x^2 + q_y^2 + q_z^2)^{1/2}$, the spatial frequency in the radial direction, is given by $q = k / 2\pi$. This then corresponds to reciprocal space, as extensively used in diffraction crystallography, while $k$-space is obtained by a simple geometric scaling. Using this definition results in a symmetric transform pair, and this symmetry can be usefully exploited in derivation of some transforms.

For the 3D case,

$$F(q) = \frac{2}{q} \int_0^\infty f(r) \sin(2\pi q r) r \, dr, \quad q > 0, \tag{12}$$

with $r = |\mathbf{r}| = \left(x^2 + y^2 + z^2\right)^{1/2}, \mathbf{r} \in \mathbb{R}^3$, where it is expressed in terms of a single-sided sine transform of $[r f(r)]$. The inverse transform is of identical form:

$$f(r) = \frac{2}{r} \int_0^\infty F(q) \sin(2\pi q r) q \, dq, \quad r > 0. \tag{13}$$

The name for this transform is not universal, but it has been called the spherically symmetric Fourier transform (here we abbreviate to SSFT) or the spherical Hankel transform. The kernel is real. The integration is along the positive real axis. The integral always delivers an even result. If $f(r)$ is an even function, the integral is one half of that from $-\infty$ to $\infty$. It will be observed from Eqs. 1, 2 that the SSFT does not depend on the form of $f(r), F(q)$ for negative $r, q$. Hence functions with different negative extensions give the same SSFT. In order to make the transform pair unique, we use an even extension for negative values of $r, q$, equivalent to replacing $r, q$ with $|r|, |q|$, which is consistent with our definition of the transform pair in Eqs. 12, 13. The standard tables of single-sided sine transforms are those of Erdelyi (Erdélyi 1954). Bracewell gives a table of only a few SSFTs (Bracewell 1978). Other tables include (Oberhettinger 1990; Poularikas 2000). Tables also exist on the web (Polyanin; eFunda). Many transforms can also be obtained from standard integral tables (Abramowitz and Stegun 1972; Gradshteyn and Ryzhik 1994). However, as none of these sources give a complete list of the transforms relevant to wave propagation, we give a list of relevant transforms in Appendix 1. We continue by calculating the SSFTs of two important functions.

The first is $f_H(r) = \sin(2\pi q_0 r)/r$. Then

$$F_H(q) = \frac{2}{q}\int_0^\infty \sin(2\pi q_0 r)\sin(2\pi q r)\,dr = \frac{1}{q}\int_{-\infty}^\infty \sin(2\pi q_0 r)\sin(2\pi q r)\,dr, \tag{14}$$

as the integrand is an even function. Introducing the complex forms for the trigonometric functions, we obtain

$$\begin{aligned}\frac{\sin(2\pi q_0 r)}{r} \Leftrightarrow \frac{1}{4q}\int_{-\infty}^{\infty} \Big\{ & \exp\left[i2\pi(q-q_0)r\right] + \exp\left[-i2\pi(q-q_0)r\right] \\ & - \exp\left[i2\pi(q+q_0)r\right] - \exp\left[-i2\pi(q+q_0)r\right]\Big\}\,dr \\ = \frac{1}{2q}[\delta(q-q_0) & - \delta(q+q_0)] = \delta(q^2 - q_0{}^2),\end{aligned} \tag{15}$$

using the properties of the Dirac delta function $\delta(\cdot)$ (Dirac 1981). As $q$ is positive definite, $\delta(q+1/\lambda)$ vanishes upon integration over positive $q$, so effectively

$$\frac{\sin(2\pi q_0 r)}{r} \Leftrightarrow \frac{1}{2q_0}\delta(q-q_0). \tag{16}$$

The inverse transform follows directly from substituting the delta function into Eq. 13. This transform, which corresponds to our homogeneous solution in Eq. 3, does not appear in many of the standard tables, nor does Mathematica incorporate it. But it is frequently used to describe propagating waves in diffraction crystallography (James 1948), grating theory (Kogelnik 1969; Sheppard 1976), diffraction and imaging theory (McCutchen 1964; Sheppard 1986), holographic reconstruction (Wolf 1969), and tomography (Dändliker and Weiss 1970; Devaney 1982). It has been very successful in calculation of focused fields, analysis of imaging systems, and in tomographic reconstruction. Then $\delta(q-1/\lambda)$ describes the surface of a sphere in reciprocal space or *k*-space, that represents the property that the magnitude of the wave vector is fixed at a real value $k = k_0$, where $k_0 = 2\pi/\lambda = 2\pi q_0$.

## *2.4 The McCutchen sphere and the Ewald sphere*

In optics, McCutchen introduced the generalized pupil, which is a cap of a sphere of radius equal to the focal length of a lens (McCutchen 1964). Then in the Debye approximation the amplitude in the focal region of the lens, when illuminated by a plane wave, is given by the 3D Fourier transform of the generalized pupil. The fact that the coherent transfer function (CTF) (in spatial frequency space) is a scaled version of the pupil is well known in 2D Fourier optics (Goodman 1968); so analogously in 3D, the 3D CTF is a scaled version of the 3D (generalized) pupil. A related concept is the Ewald sphere of diffraction crystallography, but more accurately the Ewald sphere represents the scattering vector rather than the wave vector, and so as a result the sphere is shifted so that it passes through the origin of reciprocal space rather than having its center at the origin. The McCutchen construction has the advantage over the Fresnel approximation that it is by nature non-paraxial. It has been shown that the concept of the generalized pupil can be extended to the finite Fresnel number case (Lin, Yuan et al. 2011; Kou, Sheppard et al. 2013; Sheppard, Lin et al. 2013), to electromagnetic focusing (Sheppard and Larkin 1997; Lin, Rodriguez-Herrera et al. 2012), and to pulsed waves (Sheppard 2002; Sheppard and Sharma 2002).

The term in Eq. 15 is a solution of the homogeneous Helmholtz equation (Eq. 3) and for this reason we call it the *homogeneous component* of the Green function. It is made up of only propagating plane waves (which we define to include combinations that give standing waves), and contains *no evanescent part*. It is divergence-free. For the complete spherical shell $\delta(q-1/\lambda)$, the propagating waves combine to give a pure standing wave structure. This is the simple physical example of focusing of a complete spherical in-going wave, which in free space subsequently expands to form an out-going wave (van der Pol 1936; Sheppard and Matthews 1987). Physically, we can recognize that a point in $\mathbf{q}$ space represents a propagating plane wave, which propagates inwards from infinity, before passing the origin and then propagating outwards towards infinity. This is not the same as an out-going wave. Two points in $\mathbf{q}$ space, diametrically opposite each other with respect to the origin, add together to produce a standing wave fringe pattern, and then integrating over such fringe patterns with different orientations gives a focused spot. At the focal point all the plane waves add in phase to produce a maximum in amplitude. This case corresponds to a source in conjunction with a sink. A sink we usually assume cannot exist alone in the real world (unless produced as a result of a drain, or an outlet (Tyc and Zhang 2011)). A few recent papers seem to have missed the point that isolated sinks do not physically occur in free space. Note that although the transform is represented as an integral of a real function along the positive real axis, the result, if an analytic function, can be continued analytically to complex frequencies, or in the case of the inverse transform to complex positions. In complex source-point theory, due account need be taken of branch cuts, appropriate choice of which can produce beam-like or source-like solutions (Kaiser 2000; Sheppard 2007). A source-sink pair is the basis of a version of a complex source point model of Gaussian beams that avoids nonphysical singularities (Berry 1994; Sheppard and Saghafi 1998).

## 2.5 The inhomogeneous solution, continued

Our second important example of a function is $f_I(r) = \cos(2\pi q_0 r)/r$ . Then we have

$$F_I(q) = \frac{2}{q}\int_0^{\infty}\cos(2\pi q_0 r)\sin(2\pi q r)\,dr. \tag{17}$$

This integral can be evaluated in several different ways. One such way is to use the properties of the Heaviside step function and the signum function, $\operatorname{sgn}(\cdot)$ (Bracewell 1978). Then, introducing the complex forms for the trigonometric functions,

$$\begin{aligned} F_I(q) = \frac{1}{i4\pi q}\int_{-\infty}^{\infty}\operatorname{sgn}(r)\big\{&\exp\left[i2\pi(q-q_0)r\right] - \exp\left[-i2\pi(q-q_0)r\right] \\ &+\exp\left[i2\pi(q+q_0)r\right] - \exp\left[-i2\pi(q+q_0)r\right]\big\}\,dr, \end{aligned} \tag{18}$$

and, using the Fourier transform of the signum function and the shift theorem, we obtain

$$F_I(q) = \frac{1}{2\pi q}\left(\frac{1}{q+q_0} + \frac{1}{q-q_0}\right) = \frac{1}{\pi\left(q^2 - q_0^2\right)}. \tag{19}$$

The inverse transform can be obtained rigorously as shown by Lighthill (Lighthill 1958), giving the transform pair,

$$\frac{\cos(2\pi q_0 r)}{r} \Leftrightarrow \frac{1}{\pi\left(q^2 - q_0^2\right)}. \tag{20}$$

Again this transform is not in some standard tables. (But it is in (Erdélyi 1954), (Poularikas 2000), and (Oberhettinger 1990)). In fact it is different from the standard (complex) result for the inverse transform, obtained using contour integration by assuming $q_0^2$ has a small positive imaginary part. As the solution in Eq. 19 is obtained by Fourier transformation of the inhomogeneous Helmholtz equation we call it the *inhomogeneous* part of the Green function. Note that this terminology is different from a usual connotation of inhomogeneous waves, as being synonymous with evanescent waves. In the SSFT the kernel is real, so we would expect the transform of a real function to be also real. Further, the functions on both sides of Eq. 20 are spherically symmetric in 3D space, and hence their projections on to the $r, q$ axes are even functions. By the projection-slice theorem of Fourier transforms, neither function should give rise to a complex transform.

## 2.6 The out-going wave and causality

We can combine the results of Eqs. 15 and 19 to give the transform pair

$$\frac{\exp(i2\pi q_0 r)}{r} \Leftrightarrow \frac{1}{\pi\left(q^2 - q_0^2\right)} + i\delta\left(q^2 - q_0^2\right). \tag{21}$$

With $q_0 = 1/\lambda$, this represents an outgoing spherical wave, or source, equivalent to the Green function or impulse response. Dirac showed (Dirac 1981), that an out-going wave (or particle) must have a Fourier transform of this form in order to satisfy causality (Bracewell 1978). Causality requires that the real and imaginary parts satisfy a Hilbert transform relationship (Wiener 1930; Gabor 1946; Titchmarsh 1948; Jackson 1962), as in the Kramers-Kronig relations for dispersion. In the present paper, causality also applies in space, rather than in time, because of the single-sided form of polar coordinates. Note that Jackson uses the causal form in his treatment of dispersion, and even comments on its usefulness, but does not introduce it for the Green function in space (Jackson 1962). In mathematics this form is known as the Sokhotsky-Plemelj formula.

## 2.7 On-shell and off-shell, and Bragg diffraction

The two terms of Eq. 20 are equivalent to the off-shell and on-shell conditions, respectively, in quantum field theory. Virtual particles are allowed to be off shell. The connection between virtual photons and evanescent waves has been discussed (Lawson 1970; Stahlhofen and Nimtz 2006). Lawson describes how there are two different types of virtual photon, corresponding to real off-shell values of *k*, and complex *k*, respectively. In this Section we are concerned with the first type (Lawson 1970). According to Lawson, values $k > k_0$, i.e. when the momentum is greater than the energy associated with it, correspond to virtual photons with negative mass, and are 'spacelike'. When $k < k_0$, i.e. when the momentum is less than the energy associated with it, the virtual photons have positive mass, and are 'timelike'. The second type of virtual photon, with complex *k*, corresponds with classical evanescent waves, as in Section 3.

The on-shell part of Eq. 20 corresponds to satisfaction of the Bragg diffraction condition. If this term alone existed, there would be no off-Bragg diffraction. In grating diffraction, off-Bragg diffraction is modeled using *k*-vector closure, which assumes the diffracted wave has slightly different value of $q$, or by the so-called *β*-value method, which introduces a dephasing measure (Kogelnik 1969). In dynamical theory of diffraction, a deformation of the *k*-space sphere (dispersion surface) occurs, analogous to the splitting of energy levels in quantum theory (James 1948). The inhomogeneous part can be interpreted as describing the dispersion of free space, a resonance phenomenon associated with an impulsive source.

As $q$ is positive definite, $\delta(q+q_0)$ vanishes upon integration over positive $q$. So effectively we can write

$$\frac{\exp(i2\pi q_0 r)}{r} \Leftrightarrow \frac{1}{2\pi q_0(q-q_0)} - \frac{1}{2\pi q_0(q+q_0)} + \frac{i}{2q_0}\delta\left(q-q_0\right). \tag{22}$$

## 2.8 Regularization

In the classic physics and electromagnetic books (Jackson 1962; Morse and Feshbach 1978), the homogeneous solution is not written explicitly, but can be recovered by treating $q$ as complex and choosing an appropriate integration contour. Then Eq. 21 (or Eq. 22) can be recognized as showing explicitly the real and imaginary parts of $F(q)$. The real part is the Cauchy principal value of the function at $q = q_0$, so we should write $F_I(q) = P\left[1/\pi(q^2 - q_0^2)\right]$, where $P$ means principal part. An equivalent approach is to introduce explicitly a non-physical small imaginary part (Arnoldus 2001), giving

$$\begin{aligned} F_I(q) &= \operatorname*{Lim}_{\varepsilon\to 0^+}\left[\frac{1}{\pi(q^2 - q_0^2 - i\varepsilon)}\right] = \operatorname*{Lim}_{\varepsilon\to 0^+}\left\{\frac{(q^2 - q_0^2)}{\pi[(q^2 - q_0^2)^2 + \varepsilon^2]} + \frac{i\varepsilon}{\pi[(q^2 - q_0^2)^2 + \varepsilon^2]}\right\} \\ &= \frac{1}{\pi(q^2 - q_0^2)} + i\operatorname*{Lim}_{\varepsilon\to 0^+}\left\{\frac{\varepsilon}{\pi[(q^2 - q_0^2)^2 + \varepsilon^2]}\right\} = \frac{1}{\pi(q^2 - q_0^2)} + i\delta(q^2 - q_0^2), \end{aligned} \tag{23}$$

as before. This is a form of regularization, widely used in inverse problems, similar to that used in renormalization theory. The imaginary part of Eq. 23, the second term of Cauchy-Lorentz form, gives a delta function in the limit, and represents solely propagating waves. Evanescent waves are contained in the real part only (first term), which exhibits a singularity. Sometimes the expression for $F_I(q)$ as given in Eq. 19 is understood to include implicitly the imaginary part, i.e. $q$ (and therefore $\lambda$) is assumed complex. But as Eq. 19 exhibits a branch cut along the real axis, we prefer to take both $q$ and $q_0$ as real and non-negative. Then according to Hankel's theorem (Watson 1980) the value of the transform for real $q$ is equal to the mean of the values of the transform for $\varepsilon \to 0^+$ and $\varepsilon \to 0^-$, just as we normally assume in deriving Fourier series. The integral is then equivalent to the Cauchy principal value.

## *2.9 In-going waves*

In a similar manner, for an in-going rather than an out-going wave,

$$\frac{\exp(-i2\pi q_0 r)}{r} \Leftrightarrow \frac{1}{\pi\left(q^2 - q_0^2\right)} - i\delta\left(q^2 - q_0^2\right). \tag{24}$$

This corresponds to a sink, which as we have said, we usually assume cannot exist alone in the real world. This assumption is equivalent to the Sommerfeld radiation condition (Sommerfeld 1949; Schot 1992). But a sink can exist physically in free space in conjunction with a source, as together they can combine to give the homogeneous part alone, which exhibits no singularity in the spatial domain. We recognize that

$$\frac{1}{\pi\left(q^2-q_0^2\right)} \Leftrightarrow \frac{\cos(2\pi q_0 r)}{r} = \frac{1}{2}\left[\frac{\exp(i2\pi q_0 r)}{r} + \frac{\exp(-i2\pi q_0 r)}{r}\right], \tag{25}$$

also represents a combination of an out-going and an in-going wave, each of which includes both propagating and evanescent components.

## *2.10 Homogeneous and inhomogeneous waves, versus traveling and evanescent waves*

The homogeneous component in Eq. 5 contains only propagating components. It is a special case of the Whittaker expansion over propagating plane waves traveling in all directions, corresponding to a complete sphere in Fourier space (Whittaker 1903). This should be contrasted with the Weyl expansion, which is an expansion over a half-space, giving a hemisphere of propagating waves, and evanescent waves (Weyl 1919). The Weyl expansion for the scalar Green function was presented by Carter (Carter 1975): the evanescent field contains components for $(q_x^2+q_y^2)>1/\lambda$ only, whereas in Eq. 20 $q$ can be greater than or less than $1/\lambda$. An analytic expression for the propagating and evanescent components was given by Bertilone (Bertilone 1991; Bertilone 1991). The Green function in Eq. 21 is made up of a real inhomogeneous part and an imaginary homogeneous part. Here 'homogeneous' and 'inhomogenous' refer to corresponding forms of the Helmholtz equation. We thus distinguish between 'inhomogeneous' and 'evanescent' components. Both homogeneous and inhomogeneous parts contribute to the far field, but only the inhomogeneous component contributes to the near-field singularity (Sheppard and Aguilar 2001). Each propagating component in the Weyl expansion contains inhomogeneous, as well as homogeneous, components when considered in 3D, because for a point source a propagating plane wave component exhibits a discontinuity at the source, being out-going on both sides.

## 3. Forward and backward propagating waves

We now explore the result of introducing a specific orientation. The inhomogeneous part of the transform of the Green function $1/[\pi(q^2-q_0^2)]$ contains components for $q<1/\lambda$ in addition to those for $q>1/\lambda$, and also contains propagating as well as evanescent components. Putting

$$(q^2-q_0^2)=[q_z-(q_0^2-q_x^2-q_y^2)^{1/2}][q_z+(q_0^2-q_x^2-q_y^2)^{1/2}], \tag{26}$$

the transform of the Green function can be written in the form (Sheppard, Fatemi et al. 1995; Kou, Sheppard et al. 2013; Sheppard, Lin et al. 2013)

$$\begin{aligned} F(q_x,q_y,q_z)=&\frac{1}{2\pi q_z\left(q_z-\sqrt{q_0^2-q_x^2-q_y^2}\right)}+\frac{1}{2\pi q_z\left(q_z+\sqrt{q_0^2-q_x^2-q_y^2}\right)} \\ &+\frac{i}{2|q_z|}\left[\delta\left(q_z-\sqrt{q_0^2-q_x^2-q_y^2}\right)+\delta\left(q_z+\sqrt{q_0^2-q_x^2-q_y^2}\right)\right]. \end{aligned} \tag{27}$$

Here the homogeneous part (the third and fourth terms) is written as the sum of two unweighted *hemispheres* in reciprocal space, representing forward and backward propagating waves, respectively. The total transform has only out-going propagating components, so the inhomogeneous part (the first and second terms) must also include propagating components. We thus distinguish between forward and backward propagating waves on the one hand, and out-going and in-going waves on the other. A forward propagating wave is in-going for $z<0$, and out-going for $z>0$. Some papers have confused causality with forward propagation: of course a backward propagating wave is physically realizable, as all it needs is a reversal of the *z* coordinate.

Each of the inhomogeneous terms in Eq. 27 contain components with $q_z > 0$ and $q_z < 0$. For $(q_x^2 + q_y^2) < 1/\lambda^2$ these terms each represent both out-going and in-going propagating waves, which for $z > 0$ together with the homogeneous terms result in a purely out-going, forward-propagating field, represented by a *hemispherical* shell in reciprocal space. This spherical surface, which includes contributions from both the homogeneous and inhomogeneous parts, is therefore not the same as McCutchen's spherical surface of homogeneous waves, and in fact has a strength twice that of McCutchen's.

For a source at the origin $z = 0$, the Weyl expansion can be applied separately for the regions $z < 0$, and for $z > 0$. The propagating field is then out-going, and there is also an evanescent field in both regions. But it is incorrect to think of these components as adding together to give the total field, as the expansion for $z > 0$ gives a field for $z < 0$, and vice-versa.

For $q_x^2 + q_y^2 > 1/\lambda^2$, the inhomogeneous terms in Eq. 27 transform to

$$f_I(x,y,z) = i\iint \frac{1}{q_z} \exp\left[ i2\pi\left(q_x x + q_y y\right) - 2\pi\left(q_x^2 + q_y^2 - q_0^2\right)^{1/2} |z| \right] dq_x dq_y \tag{28}$$

and are hence equivalent to the evanescent waves of the Weyl expansion. As this is true for any choice of *z* direction, we conclude that the evanescent field is completely contained in the part of the inhomogenous field corresponding to $q > 1/\lambda$. However, this part for $q > 1/\lambda$ also includes components with $(q_x^2 + q_y^2) < 1/\lambda^2$ which contributes to the propagating field. The homogeneous terms in Eq. 27 are not, by themselves, equivalent to the (out-going) propagating components of the Weyl expansion. A fuller discussion of the significance of introducing a fixed reference direction, and calculation of fields for positive $z$ only, has been presented elsewhere (Kou, Sheppard et al. 2013; Sheppard, Lin et al. 2013).

As for $z>0$ an exponential decay is also produced by a transform of the form $\delta\left(-iq_z+(q_x^2+q_y^2-1/\lambda^2)^{1/2}\right)$, we can write

$$\begin{aligned} F(q_x,q_y,q_z) &= \frac{i}{q_z}\delta\left(q_z-\sqrt{q_0^2-q_x^2-q_y^2}\right),\ \left(q_x^2+q_y^2\right)^{1/2}<q_0, q_z>0\ \&\ z>0, \\ &= \frac{1}{-iq_z}\delta\left(-iq_z+\sqrt{q_x^2+q_y^2-q_0^2}\right), \left(q_x^2+q_y^2\right)^{1/2}>q_0, -iq_z>0\ \&\ z>0. \end{aligned} \tag{29}$$

Although the properties of delta function with complex argument may not be well established, here the argument of the delta function is $-iq_z$ which is real, and we interpret the result as integration over a rectangular hyperboloid of two sheets in a spatial frequency space $(q_x,q_y,-iq_z)$. Thus the full Rayleigh-Sommerfeld diffraction formula can be applied exactly, including evanescent waves, using 3D Fourier transforms, and evaluated numerically.

## *4. Combining the Whittaker and Weyl expansions*

We consider a few different forms of expansion.

### 1. A region of space within a closed surface free of net sources.

Neglecting an evanescent field near to the outside surface, inside the region there are only propagating waves, equivalent to a sum over coincident source/sink combinations. This corresponds to a Whittaker expansion, a sum over out-going and in-going propagating components, equivalent to a divergence-free combination of sources and sinks.

### 2. A source-free half-space with forward propagating waves evanescent waves.

This corresponds to a Weyl expansion.

## 3. A finite region of space with net sources.

We can combine the Weyl and Whittaker expansions by summing over a distribution of sources and sinks with non-zero divergence, giving a total homogeneous and inhomogeneous field. The homogeneous and inhomogeneous parts of the spatial frequency content are given by the filtered transform of the Green function, given by homogeneous and inhomogeneous components. As the homogeneous part is non-zero only on the *k*-space sphere, different source/sink distributions can give rise to the same homogeneous field, but the source/sink distribution is uniquely determined by the inhomogeneous part. Relating the inhomogeneous part to the evanescent field requires selection of a reference direction *z*.

## 4. For a region $z > 0$ that is source free, any field can be generated by a sum of propagating and evanescent components.

In addition to the components in the Weyl expansion, the propagating components in general include in-going fields equivalent to sources at infinity, which give no evanescent field for finite values of *z*. For a known field in the plane $z = 0$, the spectral distribution is a function of $q_x, q_y$ only, which filters the transform corresponding to the Rayleigh-Sommerfeld kernel (Kou, Sheppard et al. 2013; Sheppard, Lin et al. 2013). For any distribution of sources and sinks with $z < 0$, the field for $z > 0$ can then be represented by an equivalent field in the plane $z = 0$. Note that we do not claim that the field in the region $z < 0$ can be determined in general by this approach. The Rayleigh hypothesis states that the field inside a re-entrant surface can be determined by analytic continuation. Recent numerical evidence and arguments based on transformation optics seem to support the hypothesis except in some pathological cases (Tishchenko 2009; Tishchenko 2010). Devaney and co-workers have attempted to reconcile the Weyl and Whittaker expansions, with some, but limited, success (Devaney and Sherman 1973; Devaney and Wolf 1974).

## 5. A list of spherically-symmetric Fourier transforms

A list of SSFTs is given in Appendix 1, based on the transforms given in Eqs. 16 and 20, rather than Eq. 8, as is taken by Mathematica. We have introduced the auxiliary functions of the sine and cosine integrals, $f_1(x) = \mathrm{Ci}(x)\sin x - \mathrm{si}(x)\cos(x)$, where $\mathrm{si}(x) = \mathrm{Si}(x) - \pi/2$. $f_1(x)$ decreases monotonically from $\pi/2$. The first four entries in the table are well known. Nos. 8-11 we have already discussed. Then Nos. 5-7 follow from 1-4 and 9. Nos. 12-14 can be derived consistently from Nos. 1-11. Some of these transforms can be generated directly using Mathematica's FourierSinTransform function. Important exceptions are Nos. 8-11, and transforms derived from them, which also do not appear in many published tables of transforms. Cases in which care is needed in applying Mathematica are indicated by an asterisk.

## 6. Summary

In summary, in contrast to the standard approach in theoretical physics and electromagnetism, which uses the transform of the Green function only as a step on the way to finding the Green function in real space, our aim is to develop the transform in its own right. The transform is written in a form where the homogeneous part is shown explicitly as a delta function representing a spherical shell, and the inhomogeneous part contains spatial frequencies with $q < q_0$ as well as $q > q_0$. Both homogeneous and inhomogeneous parts include propagating components, and the singularity of the Green function is contained completely in the inhomogeneous part for $q > q_0$.

If a particular direction of propagation is assumed, for $z > 0$ the transform consists of a *hemispherical* shell representing forward-propagating waves and part of a rectangular hyperboloid of two sheets for imaginary $q_z$ representing forward-directed evanescent waves. An analogous behavior holds for $z<0$, but the two hemispheres cannot be considered together as equivalent to the homogeneous sphere because the hemisphere for $z>0$ also gives a field for $z<0$, and vice-versa. The interpretation helps us appreciate the connection between the forward scattering model, and the use of the Ewald sphere approach in crystallographic and tomographic reconstruction.

## Appendix: Spherically-symmetric Fourier transform pairs in 3D

$$f(r)=\frac{2}{r}\int_0^\infty F(q)\sin(2\pi qr)q\,dq, r>0 \qquad F(q)=\frac{2}{q}\int_0^\infty f(r)\sin(2\pi qr)r\,dr, q>0$$

| | | |
|---|---|---|
| 1. | $\frac{1}{\pi r}$ | $\frac{1}{\pi^2 q^2}$ |
| 2. | $\frac{1}{\pi^2 r^2}$ | $\frac{1}{\pi q}$ |
| 3. | $\frac{\exp(-2\pi q_0 r)}{r}$ | $\frac{1}{\pi(q^2+q_0^2)}$ |
| 4. | $\frac{2f_1(2\pi q_0 r)}{\pi r}$ | $\frac{1}{\pi q(q+q_0)}$ |
| 5. | $\frac{2\pi\cos(2\pi q_0 r)-2f_1(2\pi q_0 r)}{\pi r}$ | $\frac{1}{\pi q(q-q_0)}$ |
| 6. | $\frac{1-2\pi ar\,f_1(2\pi q_0 r)}{2\pi^2 q_0 r^2}$ | $\frac{1-2\pi q_0 r\,f_1(2\pi q_0 r)}{2\pi^2 q_0 r^2}$ |
| 7. | $\frac{1+2\pi^2 q_0 r\cos(2\pi q_0 r)-2\pi q_0 r\,f_1(2\pi q_0 r)}{2\pi^2 q_0 r^2}$ | $\frac{1}{2\pi q_0(q-q_0)}$ |
| 8.* | $\frac{\sin(2\pi q_0 r)}{r}$ | $\frac{1}{2a}\delta(q-q_0)$ |
| 9.* | $\frac{\cos(2\pi q_0 r)}{r}$ | $\frac{1}{\pi(q^2-q_0^2)}$ |
| 10.* | $\frac{\exp(i2\pi q_0 r)}{r}$ | $\frac{1}{\pi(q^2-q_0^2)}+\frac{i}{2q_0}\delta(q-q_0)$ |
| 11.* | $\frac{\exp(-i2\pi q_0 r)}{r}$ | $\frac{1}{\pi(q^2-q_0^2)}-\frac{i}{2q_0}\delta(q-q_0)$ |
| 12.* | $\frac{1+\pi^2 q_0 r\cos(2\pi q_0 r)-2\pi q_0 r\, f_1(2\pi q_0 r)}{2\pi^2 q_0 r^2}$ | $\frac{q}{2\pi q_0(q^2-q_0^2)}$ |
| 13.* | $\frac{\pi\cos(2\pi q_0 r)-2 f_1(2\pi q_0 r)}{\pi r}$ | $\frac{a}{\pi q(q^2-q_0^2)}$ |
| 14.* | $\frac{\cos(2\pi q_0 r)-\exp(-2\pi q_0 r)}{r}$ | $\frac{2a^2}{\pi(q^4-q_0^4)}$ |

## Acknowledgments

C. J. R. Sheppard thanks the University of Melbourne for a Lyle Fellowship. S. S. Kou and J. Lin are recipients of the Discovery Early Career Researcher Award funded by the Australian Research Council under projects DE120102352 and DE130100954, respectively.

## References

Abramowitz, M. and Stegun, I. (1972). *Handbook of Mathematical Functions*. New York, Dover.

Arnoldus, H. (2001). "Representation of the near-field, middle-field, and far-field electromagnetic Green's functions in reciprocal space." *Journal of the Optical Society of America B* 18: 547-555.

Berry, M. V. (1994). "Evanescent and real waves in quantum billiards and Gaussian beams." *Journal of Physics A* 27: L391-L398.

Bertilone, D. C. (1991). "The contributions of homogeneous and evanescent plane waves to the scalar optical field : exact diffraction formulae." *J. mod. Optics* 38: 865-875.

Bertilone, D. C. (1991). "Wave theory for a converging spherical incident wave in an infinite-aperture system." *J. mod. Optics* 38: 1531-1536.

Bracewell, R. N. (1978). *The Fourier Transform and its Applications*. New York, McGraw Hill.

Carter, W. H. (1975). "Band-limited angular spectrum approximating to a spherical wave field." *J. Opt. Soc. Am.* 65: 1054-1058.

Dändliker, R. and Weiss, K. (1970). "Reconstruction of the three-dimensional refractive index from scattered waves." *Optics Comm.* 1: 323-328.

Devaney, A. J. (1982). "Inversion formula for inverse scattering within the Born approximation." *Opt. Lett.* 7: 111-113.

Devaney, A. J. and Sherman, G. C. (1973). "Plane-wave representations for scalar wave fields." *SIAM Review* 15: 765-786.

Devaney, A. J. and Wolf, E. (1974). "Multipole expansions and plane wave representations of the electromagnetic field." *Journal of Mathematical Physics* 15: 234-244.

Dirac, P. A. M. (1981). *The Principles of Quantum Mechanics*. Oxford, Oxford University Press.

eFunda (2012). http://www.efunda.com/math/Fourier_transform/table.cfm?TransName=Fs.

Erdélyi, A., Ed. (1954). *Tables of Integral Transforms*. New York, McGraw Hill.

Gabor, D. (1946). "Theory of communication." *Journal of the IEE* 93: 429-457.

Goodman, J. W. (1968). *Introduction to Fourier Optics*. New York, McGraw-Hill.

Gradshteyn, I. S. and Ryzhik, I. M. (1994). *Tables of Integrals, Series, and Products*. New York, Academic Press.

Jackson, J. D. (1962). *Classical Electrodynamics*. New York, John Wiley & Sons.

James, R. W. (1948). *The Optical Principles of the Diffraction of X-rays*. London, Bell.

Kaiser, G. (2000). Complex-distance potential theory and hyperbolic equations. *Clifford Analysis*. J. R. a. W. Sprössig. Boston, Birkhäuser.

Kogelnik, H. (1969). "Coupled wave theory for thick hologram grating." *Bell System Tech. J.* 48: 2909-2947.

Kou, S. S., Sheppard, C. J. R. and Lin, J. (2013). "Evaluation of the Rayleigh-Sommerfeld diffraction formula with 3D convolution: the 3D angular spectrum (3D-AS) method." *Opt. Lett.* 38:5296-5298

Lawson, J. D. (1970). "Some attributes of real and virtual photons." *Contemporary Physics* 11: 575-580.

Lighthill, M. J. (1958). *Introduction to Fourier Analysis and Generalised Functions*. Cambridge, Cambridge University Press.

Lin, J., Rodriguez-Herrera, O. G., Kenny, F., Lara, D. and Dainty, J. C. (2012). "Fast vectorial calculation of the volumetric focused field distribution by using a three- dimensional Fourier transform." *Optics Express* 20: 1060-1069.

Lin, J., Yuan, X. C., Kou, S. S., Sheppard, C. J. R., Rodríguez-Herrera, O. G. and Dainty, J. C. (2011). "Direct calculation of a three-dimensional diffraction field." *Opt. Lett.* 36: 1341-1343.

McCutchen, C. W. (1964). "Generalized aperture and the three-dimensional diffraction image." *J. Opt. Soc. Am.* 54: 240-244.

Morse, P. M. and Feshbach, H. (1978). *Methods of Theoretical Physics*. New York, McGraw Hill.

Oberhettinger, F. (1990). *Tables of Fourier Transforms, and Fourier Transforms of Distributions*. Berlin, Springer.

Polyanin, A. D. (2005). http://eqworld.ipmnet.ru/en/auxiliary/aux-inttrans.htm.

Poularikas, A. D. (2000). *The Transforms and Applications Handbook*. Boka Raton, CRC Press.

Schmalz, J. A., Schmalz, G., Gureyev, T. E. and Pavlov, K. M. (2010). "On the derivation of the Green's function for the Helmholtz equation using generalized functions." *American Journal of Physics* 78: 181-186.

Schot, S. (1992). "Eighty years of Sommerfeld's radiation condition." *Historia Mathematica* 19: 385-401.

Sheppard, C. J. R. (1976). "The application of the dynamical theory of x-ray diffraction to thick hologram gratings." *Int. J. Electronics* 41: 365-373.

Sheppard, C. J. R. (1986). "The spatial frequency cut-off in three-dimensional imaging." *Optik* 72: 131-133.

Sheppard, C. J. R. (2002). "Generalized Bessel pulse beams." *Journal of the Optical Society of America A* 19: 2218-2222.

Sheppard, C. J. R. (2007). "High-aperture beams: reply to comment." *Journal of the Optical Society of America A* 24: 1211-1213.

Sheppard, C. J. R. and Aguilar, J. F. (2001). "Evanescent fields do contribute to the far field (J. Mod. Opt. 1999 Vol. 46, 729) - comment." *J. mod. Optics* 48: 177-180.

Sheppard, C. J. R., Fatemi, H. and Gu, M. (1995). "The Fourier optics of near-field microscopy." *Scanning* 17: 28-40.

Sheppard, C. J. R. and Larkin, K. G. (1997). "Vectorial pupil functions and vectorial transfer functions." *Optik* 107: 79-87.

Sheppard, C. J. R., Lin, J. and Kou, S. S. (2013). "Rayleigh–Sommerfeld diffraction formula in *k* space." *Journal of the Optical Society of America A* 30: 1180-1183.

Sheppard, C. J. R. and Matthews, H. J. (1987). "Imaging in high aperture optical systems." *Journal of the Optical Society of America A* 4: 1354-1360.

Sheppard, C. J. R. and Saghafi, S. (1998). "Beam modes beyond the paraxial approximation: A scalar treatment." *Physical Review A* 57: 2971-2979.

Sheppard, C. J. R. and Sharma, M. D. (2002). "Spatial frequency content of ultrashort pulsed beams." *Journal of Optics A: Pure and Applied Optics* 4: 549-552.

Sommerfeld, A. (1949). *Partial Differential Equations in Physics*. New York, Academic Press.

Stahlhofen, A. A. and Nimtz, G. (2006). "Evanescent waves are virtual photons." *Europhysics Letters* 76: 189.

Temple, G. (1955). "The theory of generalized functions." *Proceedings of the Royal Society of London A* 228: 175-190.

Tishchenko, A. V. (2009). "Numerical demonstration of the validity of the Rayleigh hypothesis." *Optics Express* 17: 17102-17117.

Tishchenko, A. V. (2010). "Rayleigh was right: electromagnetic fields and corrugated interfaces." *Optics & Photonics News* 21: 51-54.

Titchmarsh, E. C. (1948). *Introduction to the Theory of Fourier Integrals*. Oxford, Clarendon Press.

Tyc, T. and Zhang, X. (2011). "Perfect lenses in focus." *Nature* 480: 42-43.

van der Pol, B. (1936). "On potential and wave functions in n dimensions." *Physica* 3: 385-392.

Watson, G. N. (1980). *A treatise on the theory of Bessel functions*, Cambridge University Press, Cambridge.

Weyl, H. (1919). "Ausbreitung elektromagnetische Wellen über einem ebenen Leiter." *Annalen der Physik* 365: 481-500.

Whittaker, E. T. (1903). "On the partial differential equations of mathematical physics." *Mathematische Annalen* 57: 333-355.

Wiener, N. (1930). "Generalized harmonic analysis." *Acta Mathematica* 55: 117-258.

Wolf, E. (1969). "Three-dimensional structure determination of semi-transparent objects from holographic data." *Optics Comm*. 1: 153-156.